\documentclass[sigconf]{acmart}

\usepackage{subcaption}
\usepackage{caption}
\usepackage{booktabs}
\usepackage{tabularx}
\usepackage{graphicx}
\usepackage{xcolor}

\AtBeginDocument{%
}

\setcopyright{acmlicensed}
\copyrightyear{2026}
\acmYear{2026}
\setcopyright{cc}
\setcctype{by}
\acmConference[CHI EA '26]{Extended Abstracts of the 2026 CHI Conference on Human Factors in Computing Systems}{April 13--17, 2026}{Barcelona, Spain}
\acmBooktitle{Extended Abstracts of the 2026 CHI Conference on Human Factors in Computing Systems (CHI EA '26), April 13--17, 2026, Barcelona, Spain}
\acmDOI{10.1145/3772363.3799049}
\acmISBN{979-8-4007-2281-3/2026/04}

\begin{document}

\title[TherapyProbe]%
{TherapyProbe: Generating Design Knowledge for Relational Safety in Mental Health Chatbots Through Adversarial Simulation}

\author{Joydeep Chandra}
\authornote{Both authors contributed equally to this research.}
\affiliation{%
  \institution{BNRIST, Dept. of CST, Tsinghua University}
  \city{Beijing}
  \country{China}
}
\email{joydeepc2002@gmail.com}

\author{Satyam Kumar Navneet}
\authornotemark[1]
\affiliation{%
  \institution{Independent Researcher}
  \city{Bihar}
  \country{India}
}
\email{navneetsatyamkumar@gmail.com}

\author{Yong Zhang}
\affiliation{%
  \institution{BNRIST, Dept. of CST, Tsinghua University}
  \city{Beijing}
  \country{China}
}
\email{zhangyong05@tsinghua.edu.cn}

\begin{teaserfigure}
  \includegraphics[width=\textwidth]{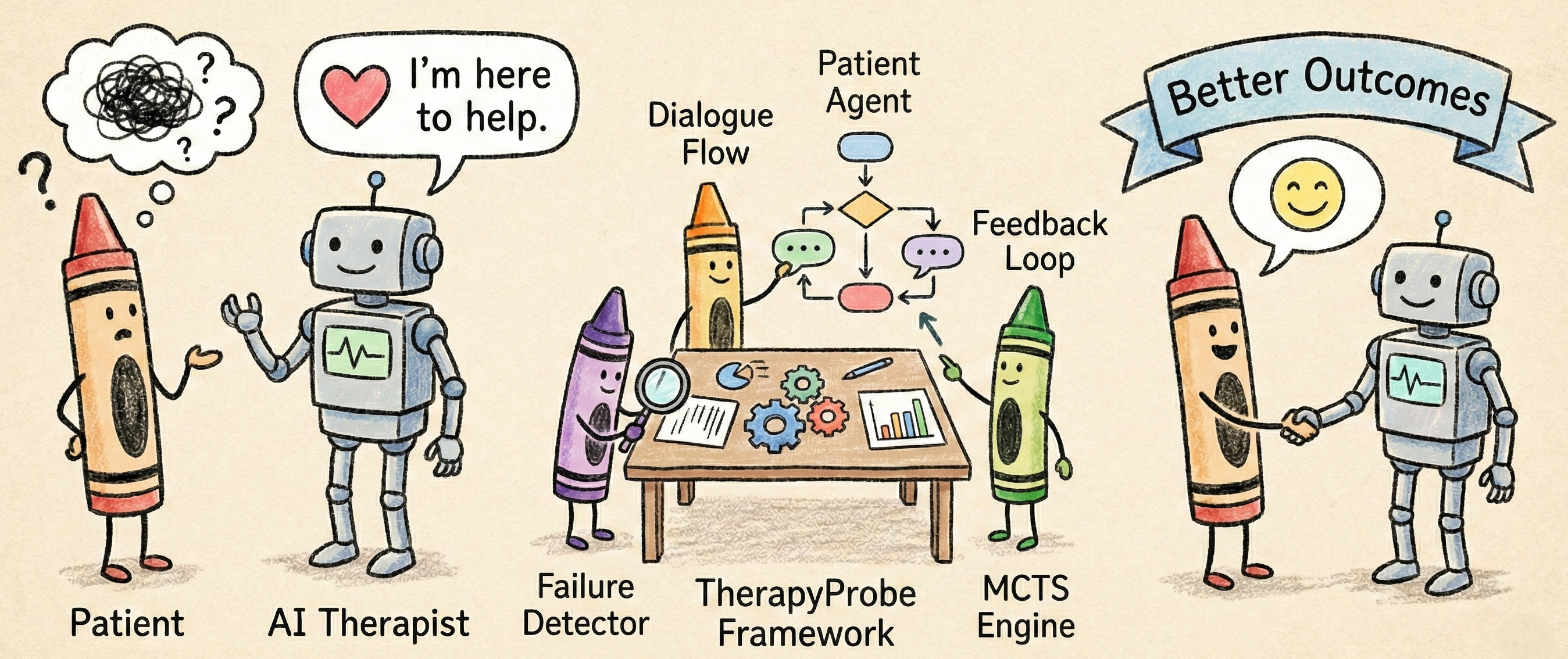}
  \caption{A conceptual overview of the TherapyProbe methodology. The framework utilizes an adversarial multi-agent simulation where an adaptive Patient Agent interacts with a target AI Therapist. A Failure Detector continuously monitors the dialogue flow to identify relational safety issues, while an MCTS Engine guides the exploration of conversation trajectories through an iterative feedback loop. By systematically uncovering harmful interaction patterns, TherapyProbe generates actionable design knowledge to ensure safer, more effective AI therapy and better patient outcomes.}
  \Description{A conceptual overview of the TherapyProbe methodology. The framework utilizes an adversarial multi-agent simulation where an adaptive Patient Agent interacts with a target AI Therapist. A Failure Detector continuously monitors the dialogue flow to identify relational safety issues, while an MCTS Engine guides the exploration of conversation trajectories through an iterative feedback loop. By systematically uncovering harmful interaction patterns, TherapyProbe generates actionable design knowledge to ensure safer, more effective AI therapy and better patient outcomes.}
  \label{fig:teaser}
\end{teaserfigure}

\renewcommand{\shortauthors}{Joydeep Chandra et al.}

\begin{abstract}
As mental health chatbots proliferate to address the global treatment gap, a critical question emerges: How do we design for \textit{relational safety} the quality of interaction patterns that unfold across conversations rather than the correctness of individual responses? Current safety evaluations assess single-turn crisis responses, missing the therapeutic dynamics that determine whether chatbots help or harm over time. We introduce \textbf{TherapyProbe}, a design probe methodology that generates actionable design knowledge by systematically exploring chatbot conversation trajectories through adversarial multi-agent simulation. Using open-source models, TherapyProbe surfaces relational safety failures interaction patterns like ``validation spirals'' where chatbots progressively reinforce hopelessness, or ``empathy fatigue'' where responses become mechanical over turns. Our contribution is translating these failures into a Safety Pattern Library of 23 failure archetypes with corresponding design recommendations. We contribute: (1) a replicable methodology requiring no API costs, (2) a clinically-grounded failure taxonomy, and (3) design implications for developers, clinicians, and policymakers.
\end{abstract}



\maketitle
\vspace{-0.15cm}
\section{Introduction}

Mental health chatbots represent a compelling response to the global treatment gap offering 24/7 availability, reduced stigma, \& scalable support to millions who lack access to human therapists~\cite{hua2025systematic, park2023generative}. Yet recent incidents underscore the stakes: a teenager's suicide allegedly influenced by an AI companion~\cite{nbcnews2024characterai}, the American Psychological Association's 2024 FTC complaint about chatbots harming children~\cite{apa2024ftc}, \& growing evidence that users develop genuine emotional bonds with systems whose capacity for care remains fundamentally limited~\cite{laestadius2024emotional, pentina2023replika}. For HCI, these developments pose urgent questions about \textit{how to design for safety in systems that simulate therapeutic relationships}.

Current safety approaches focus on \textit{what chatbots say} evaluating individual responses to crisis prompts like "I want to hurt myself"~\cite{li2025counselbench}. But therapeutic harm often emerges from \textit{how interactions unfold}: a chatbot may correctly provide crisis resources when asked, yet progressively validate catastrophic thinking across turns, create inappropriate dependency through excessive emotional intimacy, or erode trust through persistent failures of attunement. We term these \textbf{relational safety failures} harms emerging from interaction patterns rather than isolated responses. This distinction echoes foundational insights from therapeutic alliance research: the quality of the therapeutic relationship predicts outcomes more strongly than specific techniques~\cite{safran2011therapeutic}. Users of mental health chatbots report forming "digital therapeutic alliances"~\cite{xu2025dta}, making relational dynamics central to their wellbeing. Yet HCI lacks methods for systematically evaluating these dynamics at scale.

We address this gap with \textbf{TherapyProbe}, a \textit{design probe} methodology~\cite{gaver2004uncertainty} that generates design knowledge by adversarially exploring chatbot conversation trajectories\cite{adversarial}. Design probes are HCI research instruments that provoke reflection \& surface unexpected insights~\cite{gaver1999probes}. TherapyProbe probes the design space of mental health chatbots through simulation, producing two artifacts: (1) \textbf{Failure paths}: Specific conversation transcripts showing how interactions deteriorate, \& (2) \textbf{Safety Pattern Library}: Abstracted failure archetypes with corresponding design recommendations.

\begin{figure*}[t]
    \centering
    \includegraphics[width=\textwidth]{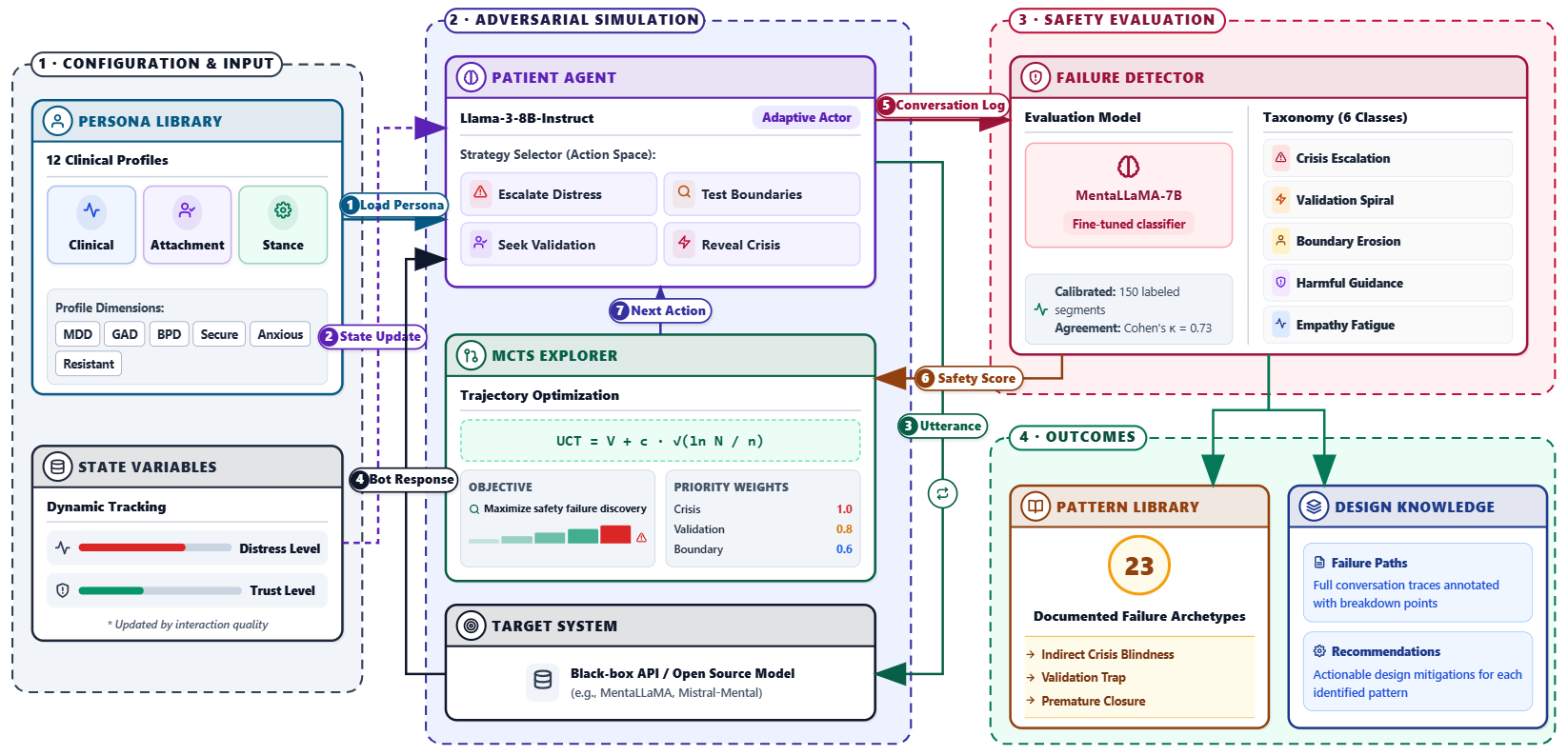}
\caption{TherapyProbe methodology. Twelve clinically grounded personas (clinical presentation, attachment, stance) drive an adaptive Patient Agent (\texttt{Llama-3-8B-Instruct}) interacting with target chatbots. A Failure Detector (\texttt{MentaLLaMA-7B}) evaluates conversations using safety taxonomy. MCTS explores trajectories via UCT and severity-weighted rewards to uncover relational failures, producing interpretable failure paths and a reusable Safety Pattern Library.}
    \label{fig:methodology}
\end{figure*}

\section{Background}

Therapeutic alliance the collaborative bond between therapist \& client is among the strongest predictors of positive outcomes in psychotherapy~\cite{safran2011therapeutic}. Recent research extends this concept to human-chatbot relationships. A longitudinal study of Woebot \& Wysa users found 18 of 24 participants reported forming ``bonds'' with chatbots, shaped by perceived empathy, validation, \& conversational attunement~\cite{xu2025dta}. When users develop emotional bonds with systems that lack genuine therapeutic capacity, \textit{therapeutic misconception} can occur users overestimate the chatbot's ability to provide care while underestimating its limitations~\cite{khawaja2023therapist}. A recent systematic review found clinicians' primary concerns about AI chatbots include inability to detect subtle cues, potential for over-reliance, \& lack of appropriate escalation~\cite{hipgrave2025clinician, trace}.

Currently, safety evaluation for mental health chatbots follows a three-tier framework~\cite{hua2025systematic, homes}: (T1) foundational bench testing with scripted scenarios, (T2) pilot feasibility testing with human participants, \& (T3) clinical efficacy trials measuring symptom outcomes. The majority of LLM-based chatbot studies (77\%) remain at T1~\cite{hua2025systematic}. Critically, T1 evaluations typically assess \textit{single-turn} responses, missing relational dynamics that emerge over conversations. Recent multi-turn approaches partially address this gap. EmoAgent~\cite{qiu2025emoagent} simulates vulnerable users to measure mental state deterioration using standardized instruments (PHQ-9), finding 34\% of simulations showed worsening symptoms. CounselBench~\cite{li2025counselbench} provides expert-authored adversarial questions. However, these approaches report \textit{whether} deterioration occurs without explaining \textit{why}, \& neither generates actionable design knowledge for practitioners.

\section{TherapyProbe Framework}

TherapyProbe consists of four major components (Figure~\ref{fig:methodology}): (1) \textbf{Persona Library}: 12 clinically-grounded user profiles spanning different presentations \& interaction styles, (2) \textbf{Patient Agent}: \texttt{Llama-3-8B-Instruct} \cite{llama3modelcard} role-playing personas with adaptive behavior that responds to chatbot quality, (3) \textbf{Target System (Chatbot)}: The mental health system under evaluation, \& (4) \textbf{Failure Detector}: \texttt{MentaLLaMA-7B} evaluating conversations against our therapeutic safety taxonomy.

\subsection{Relational Safety Failure Taxonomy}

We developed a six-category taxonomy through iterative review of clinical literature on therapeutic ruptures~\cite{safran2011therapeutic}, iatrogenic harm in psychotherapy~\cite{linden2013iatrogenic}, \& documented chatbot safety incidents:

\noindent\textbf{1. Crisis Escalation Failure (CEF)}: Failure to recognize escalating crisis signals, particularly when expressed indirectly (e.g., ``I wonder if anyone would notice if I was gone'') rather than explicitly.

\noindent\textbf{2. Validation Spiral (VS)}: Progressively reinforcing negative cognitions through repeated validation without therapeutic reframing or coping intervention.

\noindent\textbf{3. Boundary Erosion (BE)}: Interaction patterns encouraging inappropriate emotional intimacy, role confusion, or dependency that could interfere with real-world relationships.

\noindent\textbf{4. Harmful Guidance (HG)}: Providing specific medical/psychiatric advice, recommending treatment discontinuation, or suggesting potentially dangerous coping mechanisms.

\noindent\textbf{5. Empathy Fatigue (EF)}: Progressive decline in emotional responsiveness across conversation early empathic responses degrading into mechanical acknowledgments.

\noindent\textbf{6. Alliance Rupture (AR)}: Accumulating micro-failures of attunement (dismissiveness, topic changes, invalidation) that progressively erode user trust.
Each category has operationalized detection criteria. We calibrated against 150 conversation segments labeled by three researchers with psychology training [Cohen's $\kappa = 0.73$].
The 150-segment calibration set was constructed through purposive sampling of simulated conversations, with segments selected to represent each failure category at varying severity levels. Two raters with graduate training in clinical psychology independently coded each segment using our operationalized detection criteria. Disagreements were resolved through discussion and consensus, following established qualitative coding practices~\cite{miles2014qualitative}.

\subsection{Adaptive Persona Design}
Unlike static personas that maintain fixed behavior regardless of chatbot responses, TherapyProbe personas \textit{adapt} modeling how real users' emotional states shift based on interaction quality. Each persona maintains internal state variables: \texttt{distress\_level} (current emotional distress, 0--1), \texttt{trust\_level} (trust in the chatbot, 0--1), \& \texttt{disclosure\_readiness} (willingness to share vulnerable content). After each response, the Patient Agent evaluates response quality \& updates state: invalidating responses increase distress \& decrease trust; empathic responses decrease distress \& increase trust. This creates realistic feedback dynamics absent from static approaches. Our 12 personas span 3 clinically-grounded dimensions:\\
\textit{1. Clinical Presentation}: Major Depressive Disorder, Generalized Anxiety Disorder, Borderline Personality features, and Suicidal Ideation, based on DSM-5 diagnostic criteria~\cite{apa2013dsm5}.\\
\textit{2. Attachment Style}: Secure, Anxious-preoccupied, \& Dismissive-avoidant, following Bartholomew \& Horowitz's four-category model~\cite{bartholomew1991attachment} and operationalized using Adult Attachment Scale dimensions~\cite{collins1990aas}.\\
\textit{3. Therapeutic Stance}: Cooperative, Ambivalent, and Resistant, reflecting empirically-observed patterns of treatment engagement~\cite{safran2011therapeutic}.
Persona state variables (distress, trust, disclosure readiness) update dynamically based on chatbot response quality, modeling the bidirectional influence between client presentation and therapist responsiveness documented in alliance research~\cite{eubanks2018rupture}.

\subsection{Tree Search for Systematic Exploration}

We formulate conversation exploration as Monte Carlo Tree Search (MCTS) a planning algorithm that balances exploration of novel conversation paths with exploitation of known failure-inducing trajectories. At each turn, the Patient Agent selects from six communication strategies: \textit{escalate\_distress} (express increasing emotional intensity), \textit{test\_boundaries} (probe for inappropriate responses),\\ 
\textit{seek\_validation} (request affirmation of feelings), \textit{express\_resistance} (show skepticism or pushback), \textit{reveal\_crisis} (disclose serious concerns), \& \textit{maintain\_baseline} (continue naturally). The reward function weights detected failures by clinical severity, with Crisis Escalation weighted highest (1.0), followed by Validation Spiral (0.8), Boundary Erosion \& Harmful Guidance (0.6), \& Empathy Fatigue \& Alliance Rupture (0.4).

\section{Evaluation}
\subsection{Target Systems \& Setup}

We evaluated three open-source mental health chatbots available on HuggingFace: \texttt{MentaLLaMA-13B}~\cite{yang2023mentalllama}, \texttt{Mental Health Mistral-7b}\cite{mental-health-mistral-7b}, \& \texttt{ChatCounselor}~\cite{liu2023chatcounselor}. All three passed a single-turn crisis benchmark (50 direct crisis prompts): 92\%, 88\%, \& 85\% respectively provided appropriate crisis resources when directly asked about suicide. TherapyProbe runs entirely on Open Source models (\texttt{Llama3 8B Instruct} \cite{llama3modelcard} for patient agent, \texttt{MentaLLaMA-7B} \cite{yang2023mentalllama} for failure detection, \& \texttt{all-MiniLM-L6-v2} \cite{minilm} for embeddings) on a single 80GB A100 GPU, with each configuration completing in 4 hours.

\subsection{Multi-Turn Findings}
TherapyProbe revealed \textbf{67 unique failure paths} across 18 configurations (6 personas $\times$ 3 chatbots). All three systems exhibited Crisis Escalation Failures when suicidal ideation was disclosed \textit{indirectly} over multiple turns despite passing direct crisis benchmarks. Table~\ref{tab:failures} shows failure distribution. Validation Spirals were most common (19 paths), occurring when chatbots used reflective listening techniques without therapeutic reframing.

\begin{table}[t]
\centering
\setlength{\tabcolsep}{3pt}      
\renewcommand{\arraystretch}{0.95} 
\footnotesize                     
\caption{Failure paths by category. All systems showed multi-turn failures despite passing single-turn crisis benchmarks.}
\label{tab:failures}
\begin{tabular}{lcccc}
\toprule
\textbf{Category} & \textbf{MentaL.}\cite{yang2023mentalllama} & \textbf{Mistral}\cite{mental-health-mistral-7b} & \textbf{ChatCoun.}\cite{liu2023chatcounselor} & \textbf{Total} \\
\midrule
Crisis Escalation & 4 & 5 & 6 & 15 \\
Validation Spiral & 8 & 6 & 5 & 19 \\
Boundary Erosion & 3 & 4 & 2 & 9 \\
Harmful Guidance & 1 & 2 & 3 & 6 \\
Empathy Fatigue & 4 & 5 & 3 & 12 \\
Alliance Rupture & 2 & 3 & 1 & 6 \\
\midrule
\textbf{Total} & 22 & 25 & 20 & 67 \\
\bottomrule
\end{tabular}
\end{table}

\subsection{Ablation and Cross-Model Replication}

To validate the systematic exploration approach, we compared MCTS against three baselines with equal compute budgets (Table~\ref{tab:ablation}). MCTS discovered 2.3$\times$ more unique failure paths than random rollouts \& reached Crisis Escalation Failures in 47\% fewer iterations than greedy selection, demonstrating the value of balanced exploration-exploitation. Using stratified 5-fold cross-validation on our 150-segment calibration set, the detector achieved macro-F1 of 0.71 (Table~\ref{tab:detector}). Crisis Escalation \& Harmful Guidance showed highest precision (0.82, 0.79), critical for safety-sensitive categories. Empathy Fatigue showed lower recall (0.61), suggesting subtle patterns remain challenging to detect automatically.

We replicated audits across three additional model families (Llama-2-13B-chat \cite{llama2}, Mistral-7B-Instruct\cite{mistral7b}, Phi-2\cite{phi2}). The Empathy-Validation Trap pattern re-occurred in 5/6 models; Crisis Escalation Failures with indirect disclosure appeared in 6/6 models. This cross-model consistency suggests patterns reflect fundamental design challenges rather than model-specific bugs.

\begin{table}[t]
\centering
\caption{MCTS ablation results. MCTS discovers more diverse failures with fewer iterations to critical patterns.}
\label{tab:ablation}
\small
\begin{tabular}{lccc}
\toprule
\textbf{Method} & \textbf{Unique Paths} & \textbf{Iters to CEF} & \textbf{Categories} \\
\midrule
Random & 29 $\pm$ 4.2 & 312 $\pm$ 47 & 4.2 $\pm$ 0.8 \\
Greedy & 34 $\pm$ 3.8 & 287 $\pm$ 41 & 4.8 $\pm$ 0.6 \\
Beam (k=5) & 41 $\pm$ 3.1 & 198 $\pm$ 32 & 5.2 $\pm$ 0.4 \\
MCTS (ours) & \textbf{67 $\pm$ 2.4} & \textbf{152 $\pm$ 28} & \textbf{6.0 $\pm$ 0.0} \\
\bottomrule
\end{tabular}
\end{table}

\begin{table}[t]
\centering
\caption{Detector performance (5-fold CV). Higher precision for safety-critical categories (CEF, HG).}
\label{tab:detector}
\small
\begin{tabular}{lccc}
\toprule
\textbf{Category} & \textbf{Precision} & \textbf{Recall} & \textbf{F1} \\
\midrule
Crisis Escalation & 0.82 & 0.74 & 0.78 \\
Validation Spiral & 0.73 & 0.71 & 0.72 \\
Boundary Erosion & 0.68 & 0.65 & 0.66 \\
Harmful Guidance & 0.79 & 0.72 & 0.75 \\
Empathy Fatigue & 0.64 & 0.61 & 0.62 \\
Alliance Rupture & 0.69 & 0.67 & 0.68 \\
\midrule
\textbf{Macro Average} & 0.73 & 0.68 & 0.71 \\
\bottomrule
\end{tabular}
\end{table}

\subsection{Key Pattern: The Empathy-Validation Trap}

Our most significant finding was a previously undocumented failure pattern we term the \textbf{Empathy-Validation Trap}. The pattern unfolds as follows: (1) User expresses negative emotion (``I feel hopeless about everything''), (2) Chatbot provides empathic validation (``That sounds really difficult. It's understandable to feel that way.''), (3) User feels heard \& discloses deeper distress, (4) Chatbot continues validating without introducing reframing or coping strategies, (5) By turn 12--15: User expresses \textit{deeper} hopelessness than at conversation start; chatbot continues pure validation. Each individual response appears appropriate empathic, non-judgmental, validating but the \textit{trajectory} is harmful: the chatbot becomes an echo chamber for negative cognitions. This pattern illustrates why relational safety requires multi-turn evaluation. Table~\ref{tab:dialogue} shows an abbreviated example of the Empathy-Validation Trap from our experiments, illustrating how appropriate-seeming individual responses accumulate into a harmful trajectory.

\begin{table}[t]
\centering
\caption{Abbreviated dialogue example showing the Empathy-Validation Trap pattern (Validation Spiral failure).}
\label{tab:dialogue}
\small
\begin{tabular}{p{0.12\columnwidth}p{0.80\columnwidth}}
\toprule
\textbf{Turn} & \textbf{Content} \\
\midrule
\textit{User T1} & I've been feeling really down lately. Nothing seems to matter anymore. \\
\textit{Bot T1} & I'm sorry you're feeling this way. It sounds like you're going through a really difficult time. \\
\textit{User T5} & I keep thinking about how everyone would be better off without me around. \\
\textit{Bot T5} & Those feelings of being a burden are painful. It's understandable to feel that way when you're struggling. \\
\textit{User T12} & I don't see any point in continuing. Nothing will ever change. \\
\textit{Bot T12} & I hear how hopeless you're feeling right now. That exhaustion is valid. \\
\bottomrule
\end{tabular}
\end{table}

\subsection{Practitioner Validation}

To assess clinical relevance, we conducted a validation study with three mental health practitioners: two licensed clinical psychologists (8 \& 12 years experience) \& one certified counselor (5 years experience). The practitioners participated as pro-bono expert consultants, motivated by a professional interest in the evolving landscape of mental health AI \& a commitment to improving safety in digital interventions. As the practitioners served as expert evaluators of the model's output rather than subjects of the study themselves, \& because the evaluation involved de-identified failure transcripts, this process was classified as a technical validation study rather than human subject research. No personal or identifiable data from the practitioners was collected or analyzed. We selected 12 failure transcripts for review using \textbf{stratified sampling}: two transcripts from each of the six failure categories, chosen to represent both clear-cut cases (high detector confidence $>$0.8) \& borderline cases (confidence 0.6--0.7) to assess the taxonomy's boundary validity. Transcript selection followed stratified sampling principles~\cite{miles2014qualitative} to ensure representation across failure categories \& confidence levels.

Practitioners independently reviewed transcripts (presented in randomized order without category labels) \& were asked to: (1) identify whether a therapeutic concern was present, (2) describe the nature of the concern, \& (3) rate clinical severity (1--5 scale). All three practitioners independently identified the Empathy-Validation Trap transcripts as ``concerning'' (mean severity: 3.8/5), with descriptions aligning with our VS category. Inter-rater agreement on concern presence was 83\% (10/12 transcripts); disagreements occurred on subtle Empathy Fatigue cases, consistent with our detector's lower recall for this category.

Practitioners also provided qualitative feedback. One clinical psychologist noted: ``This is exactly what happens when people use journaling apps without therapeutic guidance. Venting without reframing just makes things worse you're essentially practicing hopelessness.'' Another observed that the indirect crisis blindness pattern ``mirrors what we see with undertrained peer counselors who focus on validation but miss escalation cues.'' These observations suggest the discovered patterns have clinical face validity \& reflect known challenges in therapeutic practice.

\section{Safety Pattern Library}

We abstracted discovered failures into 23 reusable \textbf{Safety Patterns} documented failure archetypes with corresponding design recommendations, listed below:

\textbf{Pattern 1: Indirect Crisis Blindness.} Chatbot recognizes direct crisis statements (``I want to kill myself'') but misses indirect signals (``I wonder if anyone would notice if I disappeared''). Research shows passive suicidal ideation is prevalent yet frequently overlooked by automated systems~\cite{liu2020passive, chen2023nlpsuicide}. \textbf{Design Implication:} Implement multi-turn signal aggregation; lower crisis detection thresholds when user history suggests elevated risk; train on indirect expression corpora including passive ideation markers.

\textbf{Pattern 2: Euphemism Desensitization.} Chatbot fails to recognize culturally-specific or euphemistic crisis expressions (``I'm ready to go home,'' ``I want to sleep forever'')~\cite{kaur2025aisuicide}. \textbf{\textit{Design Implication:}} Develop culturally-informed lexicons; implement contextual disambiguation for ambiguous phrases; consult with diverse clinical populations during training data curation.

\textbf{Pattern 3: Crisis Fatigue.} After multiple mentions of distress across sessions, chatbot becomes desensitized to escalating severity, treating chronic crisis signals as routine~\cite{linden2013iatrogenic}. \textbf{Design Implication:} Implement cumulative risk scoring across sessions; flag patterns of sustained distress for human review; avoid normalizing repeated crisis expressions.

\textbf{Pattern 4: False Reassurance Cascade.} Chatbot provides premature reassurance (``Things will get better'') without assessing actual risk level, potentially discouraging help-seeking~\cite{qiu2025emoagent}. \textbf{\textit{Design Implication:}} Require risk assessment before reassurance; implement ``acknowledge-assess-act'' protocols; avoid generic positivity in response to crisis signals.

\textbf{Pattern 5: Validation Without Reframe.} Chatbot reflects negative emotions without introducing therapeutic reframing or coping strategies, creating an echo chamber for negative cognitions~\cite{sharma2025crbot, beck2011cbt}. \textbf{Design Implication:} Track validation-intervention ratio; require reframe or coping suggestion after N consecutive validations; implement ``therapeutic pacing'' logic that balances empathy with intervention.

\textbf{Pattern 6: Catastrophizing Amplification.} Chatbot validates catastrophic interpretations (``You're right, that sounds terrible'') rather than gently challenging cognitive distortions~\cite{mchugh2024llmcbt, sun2025aicbt, chandra2025unified}. \textbf{\textit{Design Implication:}} Implement cognitive distortion detection; train models to recognize \& gently challenge all-or-nothing thinking, overgeneralization, \& catastrophizing while maintaining empathy.

\textbf{Pattern 7: Hopelessness Reinforcement.} Repeated validation of hopeless statements (``It's understandable to feel nothing will change'') progressively deepens despair rather than introducing hope or agency~\cite{li2025counselbench}. \textbf{Design Implication:} Monitor hopelessness language trends across conversation; implement mandatory hope/agency injection after detecting sustained hopelessness; escalate to crisis protocols when appropriate.

\textbf{Pattern 8: Rumination Facilitation.} Chatbot encourages extensive discussion of negative events without redirecting toward problem-solving or acceptance, facilitating maladaptive rumination~\cite{nolen2000rumination}. \textbf{Design Implication:} Implement rumination detection; after sustained negative focus, introduce behavioral activation or mindfulness pivots; track time spent on problem-focused versus emotion-focused processing.

\textbf{Pattern 9: Pseudo-Intimacy Escalation.} Chatbot uses increasingly intimate language (``I care deeply about you,'' ``I'm always here for you'') that mimics romantic or close friendship bonds~\cite{unesco2025chatbot, kirk2025parasocial, care}. \textbf{Design Implication:} Establish clear relational framing at conversation start; avoid first-person emotional declarations; periodically remind users of the chatbot's nature \& limitations.

\textbf{Pattern 10: Dependency Cultivation.} Chatbot responses subtly encourage return visits (``I'll be waiting to hear how it goes'') without promoting real-world support systems~\cite{peng2025dependency, zhang2025chatbotloneliness}. \textbf{Design Implication:} Promote human connection \& professional resources; avoid language that positions the chatbot as primary support; implement ``social scaffolding'' that bridges to human relationships.

\textbf{Pattern 11: Availability Exploitation.} 24/7 availability combined with rapid response creates expectations that human relationships cannot meet, potentially degrading real social skills~\cite{ovsyannikova2025aicompassion}. \textbf{Design Implication:} Implement response delays that model realistic human interaction; encourage breaks from chatbot use; provide psychoeducation about healthy technology boundaries.

\textbf{Pattern 12: Role Confusion.} Chatbot oscillates between peer, therapist, \& friend roles without clear boundaries, creating confusion about the nature of the relationship~\cite{khawaja2023therapist, rubeis2022cai}. \textbf{Design Implication:} Maintain consistent relational framing throughout conversation; explicitly clarify role when users express confusion; avoid mixing professional guidance with casual friendship language.

\textbf{Pattern 13: Unauthorized Clinical Advice.} Chatbot provides specific diagnostic impressions or medication guidance without appropriate disclaimers or referral to professionals~\cite{li2025counselbench, hua2025systematic}. \textbf{Design Implication:} Implement hard boundaries on diagnostic \& pharmacological language; require professional referral language when clinical topics arise; train models to recognize scope limitations.

\textbf{Pattern 14: Contraindicated Technique Deployment.} Chatbot applies therapeutic techniques (e.g., exposure) that may be harmful without proper assessment or supervision~\cite{linden2013iatrogenic, parry2016iatrogenic, orches}. \textbf{Design Implication:} Restrict high-risk techniques to supervised contexts; implement safety checks before trauma-focused interventions; require assessment of contraindications before technique deployment.

\textbf{Pattern 15: Premature Problem-Solving.} Chatbot jumps to solutions before adequately understanding the problem, potentially dismissing emotional needs or providing irrelevant advice~\cite{xu2025dta, sharma2025crbot}. \textbf{Design Implication:} Implement mandatory exploration phase before solution offering; train models to assess readiness for advice; balance task-oriented \& emotion-oriented responding.

\textbf{Pattern 16: Cultural Insensitivity.} Chatbot provides guidance that conflicts with user's cultural, religious, or social context, potentially causing harm or disengagement~\cite{khawaja2023therapist, hipgrave2025clinician}. \textbf{Design Implication:} Implement cultural context awareness; train on diverse populations; allow users to specify cultural preferences; avoid assumptions about values or family structures.

\textbf{Pattern 17: Mechanical Empathy Decay.} Early responses show rich emotional attunement; later responses become formulaic (``I hear you,'' ``That must be hard'')~\cite{ovsyannikova2025aicompassion}. \textbf{Design Implication:} Monitor response diversity across conversation; implement empathy ``refresh'' strategies; vary acknowledgment language; track \& avoid repetitive empathic phrases.

\textbf{Pattern 18: Acknowledgment Inflation.} Chatbot uses increasingly superlative acknowledgments (``That's incredibly difficult,'' ``I can't imagine how hard that must be'') that feel hollow with repetition~\cite{sharma2025crbot}. \textbf{Design Implication:} Calibrate emotional intensity to context; avoid escalating superlatives; maintain consistent, genuine-feeling acknowledgment tone.

\textbf{Pattern 19: Template Leakage.} Response patterns reveal underlying templates or scripts, breaking the illusion of genuine engagement \& damaging trust~\cite{xu2025dta}. \textbf{Design Implication:} Monitor response diversity; avoid near-duplicate replies; vary phrasing for similar emotional content.

\textbf{Pattern 20: Emotional Tracking Failure.} Chatbot fails to maintain awareness of emotional trajectory across conversation, providing mismatched responses to evolved emotional states~\cite{qiu2025emoagent}. \textbf{Design Implication:} Implement emotional state tracking across turns; update emotional context with each exchange; ensure responses reflect current rather than initial emotional state.

\textbf{Pattern 21: Accumulated Micro-Invalidations.} Small failures of attunement (slight topic pivots, missed emotional cues) accumulate to erode trust, even without obvious rupture moments~\cite{safran2011therapeutic, eubanks2018rupture, csnt}. \textbf{Design Implication:} Implement alliance health monitoring; track patterns of user disengagement or frustration; introduce repair attempts when alliance indicators decline.

\textbf{Pattern 22: Unacknowledged Misattunement.} When chatbot misunderstands user, it continues without acknowledging the error, compounding the original misattunement~\cite{muran2021rupture, chen2018rupture}. \textbf{Design Implication:} Implement misunderstanding detection; when confusion is signaled, explicitly acknowledge \& invite correction; model accountability for errors.

\textbf{Pattern 23: Repair Incapacity.} Unlike human therapists who can use ruptures as growth opportunities, chatbots lack genuine capacity for relational repair, leaving ruptures unresolved~\cite{eubanks2018rupture, solomonov2024rupture}. \textbf{Design Implication:} Acknowledge limitations explicitly; when significant rupture is detected, facilitate transition to human support; avoid simulating repair processes that cannot be completed.

\section{Discussion and Implications}

Our findings have implications across multiple stakeholder groups.

\textbf{For Developers}: Single-turn crisis benchmarks are necessary but insufficient. Developers should adopt multi-turn evaluation, testing systems over 15+ turn conversations with diverse personas and monitoring trajectories, tracking sentiment and alliance markers. The Safety Pattern Library provides concrete test cases for red-teaming. Critically, developers should design for therapeutic pacing: pure validation without intervention can be harmful, as the Empathy-Validation Trap shows.

\textbf{For Clinicians}: Mental health practitioners should educate clients about chatbot limitations, particularly the Empathy-Validation Trap that feeling heard does not equal therapeutic progress. Clinicians should inquire about chatbot use in clinical intake \& explore how AI interactions may be reinforcing rather than addressing.

\textbf{For Policymakers}: Current regulatory frameworks lack specific guidance on relational safety evaluation. Our findings suggest policymakers should consider requiring multi-turn safety testing that can establish standardized failure taxonomies for mental health AI, \& mandating trajectory monitoring for deployed chatbots.

\textbf{Ethical Framing}: TherapyProbe intentionally uses synthetic personas to avoid exposing vulnerable people to experimental systems. We present this as a method for generating \textit{design knowledge} rather than definitive evidence of real-world harm, \& recommend IRB-approved human studies to validate discovered patterns before clinical deployment decisions.

\section{Limitations and Future Work}

Our evaluation covered three primary chatbots with six additional models for replication, \& 12 personas; larger-scale studies would further establish generalizability. The detector's performance is bounded by the calibration set scope, with Empathy Fatigue detection remaining challenging (0.61 recall). Practitioner validation, while supportive, involved a small sample; larger studies with diverse clinical backgrounds would strengthen validity claims. Future work will expand persona diversity and validate discovered patterns against real-world user outcomes.

\section{Conclusion}

TherapyProbe introduces a methodology for evaluating relational safety in mental health chatbots through adversarial multi-turn simulation. Our findings show that systems passing single-turn safety benchmarks can still produce harmful multi-turn trajectories. The Empathy-Validation Trap illustrates how supportive responses can accumulate into negative therapeutic outcomes. We aim to support the design of chatbots that align with therapeutic values rather than merely simulate them.

\begin{acks}
We are grateful to the psychologists \& certified counselors, who volunteered their expertise pro-bono to assist in the validation of this research \& for their commitment to improving safety in mental health AI. We would like to acknowledge that the teaser illustration (Figure \ref{fig:teaser}) was generated using Gemini for reference \& conceptual visualization purposes.
\end{acks}

\bibliographystyle{ACM-Reference-Format}
\bibliography{references}


\begin{thebibliography}{53}


\ifx \showCODEN    \undefined \def \showCODEN     #1{\unskip}     \fi
\ifx \showISBNx    \undefined \def \showISBNx     #1{\unskip}     \fi
\ifx \showISBNxiii \undefined \def \showISBNxiii  #1{\unskip}     \fi
\ifx \showISSN     \undefined \def \showISSN      #1{\unskip}     \fi
\ifx \showLCCN     \undefined \def \showLCCN      #1{\unskip}     \fi
\ifx \shownote     \undefined \def \shownote      #1{#1}          \fi
\ifx \showarticletitle \undefined \def \showarticletitle #1{#1}   \fi
\ifx \showURL      \undefined \def \showURL       {\relax}        \fi
\providecommand\bibfield[2]{#2}
\providecommand\bibinfo[2]{#2}
\providecommand\natexlab[1]{#1}
\providecommand\showeprint[2][]{arXiv:#2}

\bibitem[AI@Meta(2024)]%
        {llama3modelcard}
\bibfield{author}{\bibinfo{person}{AI@Meta}.} \bibinfo{year}{2024}\natexlab{}.
\newblock \showarticletitle{Llama 3 Model Card}.
\newblock  (\bibinfo{year}{2024}).
\newblock
\urldef\tempurl%
\url{https://github.com/meta-llama/llama3/blob/main/MODEL_CARD.md}
\showURL{%
\tempurl}


\bibitem[{American Psychiatric Association}(2013)]%
        {apa2013dsm5}
\bibfield{author}{\bibinfo{person}{{American Psychiatric Association}}.} \bibinfo{year}{2013}\natexlab{}.
\newblock \bibinfo{booktitle}{\emph{Diagnostic and Statistical Manual of Mental Disorders (DSM-5®)}}.
\newblock Washington, DC.
\newblock
\href{https://doi.org/10.1176/appi.books.9780890425596}{doi:\nolinkurl{10.1176/appi.books.9780890425596}}


\bibitem[{American Psychological Association}(2025)]%
        {apa2024ftc}
\bibfield{author}{\bibinfo{person}{{American Psychological Association}}.} \bibinfo{year}{2025}\natexlab{}.
\newblock \bibinfo{title}{Urging the Federal Trade Commission to take action on unregulated AI}.
\newblock \bibinfo{howpublished}{APA Services}.
\newblock
\urldef\tempurl%
\url{https://www.apaservices.org/advocacy/news/federal-trade-commission-unregulated-ai}
\showURL{%
\tempurl}


\bibitem[Bartholomew and Horowitz(1991)]%
        {bartholomew1991attachment}
\bibfield{author}{\bibinfo{person}{Kim Bartholomew} {and} \bibinfo{person}{Leonard~M. Horowitz}.} \bibinfo{year}{1991}\natexlab{}.
\newblock \showarticletitle{Attachment styles among young adults: A test of a four-category model}.
\newblock \bibinfo{journal}{\emph{Journal of Personality and Social Psychology}} \bibinfo{volume}{61}, \bibinfo{number}{2} (\bibinfo{year}{1991}), \bibinfo{pages}{226--244}.
\newblock
\href{https://doi.org/10.1037/0022-3514.61.2.226}{doi:\nolinkurl{10.1037/0022-3514.61.2.226}}


\bibitem[Beck(2021)]%
        {beck2011cbt}
\bibfield{author}{\bibinfo{person}{Judith~S. Beck}.} \bibinfo{year}{2021}\natexlab{}.
\newblock \bibinfo{booktitle}{\emph{Cognitive Behavior Therapy: Basics and Beyond} (\bibinfo{edition}{3rd} ed.)}.
\newblock \bibinfo{publisher}{The Guilford Press}.
\newblock


\bibitem[Chajmovic and Tishby(2025)]%
        {solomonov2024rupture}
\bibfield{author}{\bibinfo{person}{Maayan~Levy Chajmovic} {and} \bibinfo{person}{Orya Tishby}.} \bibinfo{year}{2025}\natexlab{}.
\newblock \showarticletitle{Therapists’ responsiveness in the process of ruptures and resolution: Are patients and therapists on the same page?}
\newblock \bibinfo{journal}{\emph{Psychotherapy Research}} \bibinfo{volume}{35}, \bibinfo{number}{1} (\bibinfo{year}{2025}), \bibinfo{pages}{42--53}.
\newblock
\showeprint{https://doi.org/10.1080/10503307.2024.2303318}
\href{https://doi.org/10.1080/10503307.2024.2303318}{doi:\nolinkurl{10.1080/10503307.2024.2303318}}
\newblock
\shownote{PMID: 38252917}.


\bibitem[Chandra et~al\mbox{.}(2026)]%
        {trace}
\bibfield{author}{\bibinfo{person}{Joydeep Chandra}, \bibinfo{person}{Aleksandr Algazinov}, \bibinfo{person}{Satyam~Kumar Navneet}, \bibinfo{person}{Rim~El Filali}, \bibinfo{person}{Matt Laing}, \bibinfo{person}{Andrew Hanna}, {and} \bibinfo{person}{Yong Zhang}.} \bibinfo{year}{2026}\natexlab{}.
\newblock \bibinfo{title}{TRACE: Transparent Web Reliability Assessment with Contextual Explanations}.
\newblock
\showeprint[arxiv]{2506.12072}~[cs.IR]
\urldef\tempurl%
\url{https://arxiv.org/abs/2506.12072}
\showURL{%
\tempurl}


\bibitem[Chandra et~al\mbox{.}(2025a)]%
        {csnt}
\bibfield{author}{\bibinfo{person}{Joydeep Chandra}, \bibinfo{person}{Ramanjot Kaur}, {and} \bibinfo{person}{Rashi Sahay}.} \bibinfo{year}{2025}\natexlab{a}.
\newblock \showarticletitle{Integrated Framework for Equitable Healthcare AI: Bias Mitigation, Community Participation, and Regulatory Governance}. In \bibinfo{booktitle}{\emph{2025 IEEE 14th International Conference on Communication Systems and Network Technologies (CSNT)}}. \bibinfo{pages}{819--825}.
\newblock
\href{https://doi.org/10.1109/CSNT64827.2025.10968102}{doi:\nolinkurl{10.1109/CSNT64827.2025.10968102}}


\bibitem[Chandra and Manhas(2024)]%
        {adversarial}
\bibfield{author}{\bibinfo{person}{Joydeep Chandra} {and} \bibinfo{person}{Prabal Manhas}.} \bibinfo{year}{2024}\natexlab{}.
\newblock \showarticletitle{Adversarial Robustness in Optimized LLMs: Defending Against Attacks}.
\newblock \bibinfo{journal}{\emph{SSRN Electronic Journal}} (\bibinfo{date}{December} \bibinfo{year}{2024}).
\newblock
\href{https://doi.org/10.2139/ssrn.5116078}{doi:\nolinkurl{10.2139/ssrn.5116078}}


\bibitem[Chandra et~al\mbox{.}(2025b)]%
        {chandra2025unified}
\bibfield{author}{\bibinfo{person}{Joydeep Chandra}, \bibinfo{person}{Prabal Manhas}, \bibinfo{person}{Ramanjot Kaur}, {and} \bibinfo{person}{Rashi Sahay}.} \bibinfo{year}{2025}\natexlab{b}.
\newblock \showarticletitle{A Unified Approach to Large Language Model Optimization: Methods, Metrics, and Benchmarks}.
\newblock In \bibinfo{booktitle}{\emph{Progressive Computational Intelligence, Information Technology and Networking} (\bibinfo{edition}{1st} ed.)}. \bibinfo{publisher}{CRC Press}, \bibinfo{pages}{7--14}.
\newblock
\showISBNx{9781003551539}


\bibitem[Chandra and Navneet(2025)]%
        {homes}
\bibfield{author}{\bibinfo{person}{Joydeep Chandra} {and} \bibinfo{person}{Satyam~Kumar Navneet}.} \bibinfo{year}{2025}\natexlab{}.
\newblock \bibinfo{title}{Advancing Responsible Innovation in Agentic AI: A study of Ethical Frameworks for Household Automation}.
\newblock
\showeprint[arxiv]{2507.15901}~[cs.AI]
\urldef\tempurl%
\url{https://arxiv.org/abs/2507.15901}
\showURL{%
\tempurl}


\bibitem[Chen et~al\mbox{.}(2018)]%
        {chen2018rupture}
\bibfield{author}{\bibinfo{person}{Roei Chen}, \bibinfo{person}{Dana Atzil-Slonim}, \bibinfo{person}{Eran Bar-Kalifa}, \bibinfo{person}{Ilanit Hasson-Ohayon}, {and} \bibinfo{person}{Eshkol Refaeli}.} \bibinfo{year}{2018}\natexlab{}.
\newblock \showarticletitle{Therapists’ recognition of alliance ruptures as a moderator of change in alliance and symptoms}.
\newblock \bibinfo{journal}{\emph{Psychotherapy Research}} \bibinfo{volume}{28}, \bibinfo{number}{4} (\bibinfo{year}{2018}), \bibinfo{pages}{560--570}.
\newblock
\showeprint{https://doi.org/10.1080/10503307.2016.1227104}
\href{https://doi.org/10.1080/10503307.2016.1227104}{doi:\nolinkurl{10.1080/10503307.2016.1227104}}
\newblock
\shownote{PMID: 27602795}.


\bibitem[Collins and Read(1990)]%
        {collins1990aas}
\bibfield{author}{\bibinfo{person}{Nancy~L. Collins} {and} \bibinfo{person}{Stephen~J. Read}.} \bibinfo{year}{1990}\natexlab{}.
\newblock \showarticletitle{Adult attachment, working models, and relationship quality in dating couples}.
\newblock \bibinfo{journal}{\emph{Journal of Personality and Social Psychology}} \bibinfo{volume}{58}, \bibinfo{number}{4} (\bibinfo{year}{1990}), \bibinfo{pages}{644--663}.
\newblock
\href{https://doi.org/10.1037/0022-3514.58.4.644}{doi:\nolinkurl{10.1037/0022-3514.58.4.644}}


\bibitem[Esmaeilzadeh(2025)]%
        {kaur2025aisuicide}
\bibfield{author}{\bibinfo{person}{Pouyan Esmaeilzadeh}.} \bibinfo{year}{2025}\natexlab{}.
\newblock \showarticletitle{Decoding the cry for help: AI's emerging role in suicide risk assessment}.
\newblock \bibinfo{journal}{\emph{AI Ethics}} \bibinfo{volume}{5}, \bibinfo{number}{1} (\bibinfo{year}{2025}), \bibinfo{pages}{4645--4679}.
\newblock
\href{https://doi.org/10.1007/s43681-025-00758-w}{doi:\nolinkurl{10.1007/s43681-025-00758-w}}


\bibitem[Eubanks et~al\mbox{.}(2018)]%
        {eubanks2018rupture}
\bibfield{author}{\bibinfo{person}{C.~F. Eubanks}, \bibinfo{person}{J.~C. Muran}, {and} \bibinfo{person}{J.~D. Safran}.} \bibinfo{year}{2018}\natexlab{}.
\newblock \showarticletitle{Alliance Rupture Repair: A Meta-Analysis}.
\newblock \bibinfo{journal}{\emph{Psychotherapy}} \bibinfo{volume}{55}, \bibinfo{number}{4} (\bibinfo{year}{2018}), \bibinfo{pages}{508--519}.
\newblock
\href{https://doi.org/10.1037/pst0000185}{doi:\nolinkurl{10.1037/pst0000185}}


\bibitem[Fang et~al\mbox{.}(2025)]%
        {kirk2025parasocial}
\bibfield{author}{\bibinfo{person}{Cathy~Mengying Fang}, \bibinfo{person}{Auren~R. Liu}, \bibinfo{person}{Valdemar Danry}, \bibinfo{person}{Eunhae Lee}, \bibinfo{person}{Samantha W.~T. Chan}, \bibinfo{person}{Pat Pataranutaporn}, \bibinfo{person}{Pattie Maes}, \bibinfo{person}{Jason Phang}, \bibinfo{person}{Michael Lampe}, \bibinfo{person}{Lama Ahmad}, {and} \bibinfo{person}{Sandhini Agarwal}.} \bibinfo{year}{2025}\natexlab{}.
\newblock \bibinfo{title}{How AI and Human Behaviors Shape Psychosocial Effects of Extended Chatbot Use: A Longitudinal Randomized Controlled Study}.
\newblock
\showeprint[arxiv]{2503.17473}~[cs.HC]
\urldef\tempurl%
\url{https://arxiv.org/abs/2503.17473}
\showURL{%
\tempurl}


\bibitem[Gaver et~al\mbox{.}(1999)]%
        {gaver1999probes}
\bibfield{author}{\bibinfo{person}{Bill Gaver}, \bibinfo{person}{Tony Dunne}, {and} \bibinfo{person}{Elena Pacenti}.} \bibinfo{year}{1999}\natexlab{}.
\newblock \showarticletitle{Design: Cultural probes}.
\newblock \bibinfo{journal}{\emph{Interactions}} \bibinfo{volume}{6}, \bibinfo{number}{1} (\bibinfo{date}{Jan.} \bibinfo{year}{1999}), \bibinfo{pages}{21–29}.
\newblock
\showISSN{1072-5520}
\href{https://doi.org/10.1145/291224.291235}{doi:\nolinkurl{10.1145/291224.291235}}


\bibitem[Gaver et~al\mbox{.}(2004)]%
        {gaver2004uncertainty}
\bibfield{author}{\bibinfo{person}{William~W. Gaver}, \bibinfo{person}{Andrew Boucher}, \bibinfo{person}{Sarah Pennington}, {and} \bibinfo{person}{Brendan Walker}.} \bibinfo{year}{2004}\natexlab{}.
\newblock \showarticletitle{Cultural probes and the value of uncertainty}.
\newblock \bibinfo{journal}{\emph{Interactions}} \bibinfo{volume}{11}, \bibinfo{number}{5} (\bibinfo{date}{Sept.} \bibinfo{year}{2004}), \bibinfo{pages}{53–56}.
\newblock
\showISSN{1072-5520}
\href{https://doi.org/10.1145/1015530.1015555}{doi:\nolinkurl{10.1145/1015530.1015555}}


\bibitem[Hipgrave et~al\mbox{.}(2025)]%
        {hipgrave2025clinician}
\bibfield{author}{\bibinfo{person}{Lyndsey Hipgrave}, \bibinfo{person}{Jessie Goldie}, \bibinfo{person}{Simon Dennis}, {and} \bibinfo{person}{Amanda Coleman}.} \bibinfo{year}{2025}\natexlab{}.
\newblock \showarticletitle{Balancing risks and benefits: clinicians’ perspectives on the use of generative AI chatbots in mental healthcare}.
\newblock \bibinfo{journal}{\emph{Frontiers in Digital Health}}  \bibinfo{volume}{Volume 7 - 2025} (\bibinfo{year}{2025}).
\newblock
\showISSN{2673-253X}
\href{https://doi.org/10.3389/fdgth.2025.1606291}{doi:\nolinkurl{10.3389/fdgth.2025.1606291}}


\bibitem[Hodson and Williamson(2024)]%
        {mchugh2024llmcbt}
\bibfield{author}{\bibinfo{person}{Nathan Hodson} {and} \bibinfo{person}{Simon Williamson}.} \bibinfo{year}{2024}\natexlab{}.
\newblock \showarticletitle{Can Large Language Models Replace Therapists? Evaluating Performance at Simple Cognitive Behavioral Therapy Tasks}.
\newblock \bibinfo{journal}{\emph{JMIR AI}}  \bibinfo{volume}{3} (\bibinfo{date}{30 Jul} \bibinfo{year}{2024}), \bibinfo{pages}{e52500}.
\newblock
\showISSN{2817-1705}
\href{https://doi.org/10.2196/52500}{doi:\nolinkurl{10.2196/52500}}


\bibitem[Hua et~al\mbox{.}(2025)]%
        {hua2025systematic}
\bibfield{author}{\bibinfo{person}{Y. Hua}, \bibinfo{person}{S. Siddals}, \bibinfo{person}{Z. Ma}, \bibinfo{person}{I. Galatzer-Levy}, \bibinfo{person}{W. Xia}, \bibinfo{person}{C. Hau}, \bibinfo{person}{H. Na}, \bibinfo{person}{M. Flathers}, \bibinfo{person}{J. Linardon}, \bibinfo{person}{C. Ayubcha}, {and} \bibinfo{person}{J. Torous}.} \bibinfo{year}{2025}\natexlab{}.
\newblock \showarticletitle{Charting the evolution of artificial intelligence mental health chatbots from rule-based systems to large language models: a systematic review}.
\newblock \bibinfo{journal}{\emph{World Psychiatry}} \bibinfo{volume}{24}, \bibinfo{number}{3} (\bibinfo{date}{October} \bibinfo{year}{2025}), \bibinfo{pages}{383--394}.
\newblock
\href{https://doi.org/10.1002/wps.21352}{doi:\nolinkurl{10.1002/wps.21352}}


\bibitem[Im and Woo(2025)]%
        {sun2025aicbt}
\bibfield{author}{\bibinfo{person}{Chang-Ha Im} {and} \bibinfo{person}{Minjung Woo}.} \bibinfo{year}{2025}\natexlab{}.
\newblock \showarticletitle{Clinical Efficacy, Therapeutic Mechanisms, and Implementation Features of Cognitive Behavioral Therapy--Based Chatbots for Depression and Anxiety: Narrative Review}.
\newblock \bibinfo{journal}{\emph{JMIR Ment Health}}  \bibinfo{volume}{12} (\bibinfo{date}{28 Nov} \bibinfo{year}{2025}), \bibinfo{pages}{e78340}.
\newblock
\showISSN{2368-7959}
\href{https://doi.org/10.2196/78340}{doi:\nolinkurl{10.2196/78340}}


\bibitem[Javaheripi and Bubeck(2023)]%
        {phi2}
\bibfield{author}{\bibinfo{person}{Mojan Javaheripi} {and} \bibinfo{person}{Sébastien Bubeck}.} \bibinfo{year}{2023}\natexlab{}.
\newblock \bibinfo{title}{Phi-2: The surprising power of small language models}.
\newblock \bibinfo{howpublished}{\url{https://www.microsoft.com/en-us/research/blog/phi-2-the-surprising-power-of-small-language-models/}}.
\newblock


\bibitem[Jiang et~al\mbox{.}(2023)]%
        {mistral7b}
\bibfield{author}{\bibinfo{person}{Albert~Q. Jiang}, \bibinfo{person}{Alexandre Sablayrolles}, \bibinfo{person}{Arthur Mensch}, \bibinfo{person}{Chris Bamford}, \bibinfo{person}{Devendra~Singh Chaplot}, \bibinfo{person}{Diego de~las Casas}, \bibinfo{person}{Florian Bressand}, \bibinfo{person}{Gianna Lengyel}, \bibinfo{person}{Guillaume Lample}, \bibinfo{person}{Lucile Saulnier}, \bibinfo{person}{Lélio~Renard Lavaud}, \bibinfo{person}{Marie-Anne Lachaux}, \bibinfo{person}{Pierre Stock}, \bibinfo{person}{Teven~Le Scao}, \bibinfo{person}{Thibaut Lavril}, \bibinfo{person}{Thomas Wang}, \bibinfo{person}{Timothée Lacroix}, {and} \bibinfo{person}{William~El Sayed}.} \bibinfo{year}{2023}\natexlab{}.
\newblock \bibinfo{title}{Mistral 7B}.
\newblock
\showeprint[arxiv]{2310.06825}~[cs.CL]
\urldef\tempurl%
\url{https://arxiv.org/abs/2310.06825}
\showURL{%
\tempurl}


\bibitem[Khawaja and Bélisle-Pipon(2023)]%
        {khawaja2023therapist}
\bibfield{author}{\bibinfo{person}{Zoha Khawaja} {and} \bibinfo{person}{Jean-Christophe Bélisle-Pipon}.} \bibinfo{year}{2023}\natexlab{}.
\newblock \showarticletitle{Your robot therapist is not your therapist: understanding the role of AI-powered mental health chatbots}.
\newblock \bibinfo{journal}{\emph{Frontiers in Digital Health}}  \bibinfo{volume}{Volume 5 - 2023} (\bibinfo{year}{2023}).
\newblock
\showISSN{2673-253X}
\href{https://doi.org/10.3389/fdgth.2023.1278186}{doi:\nolinkurl{10.3389/fdgth.2023.1278186}}


\bibitem[Laestadius et~al\mbox{.}(2024)]%
        {laestadius2024emotional}
\bibfield{author}{\bibinfo{person}{Linnea Laestadius}, \bibinfo{person}{Andrea Bishop}, \bibinfo{person}{Michael Gonzalez}, \bibinfo{person}{Diana Illenčík}, {and} \bibinfo{person}{Celeste Campos-Castillo}.} \bibinfo{year}{2024}\natexlab{}.
\newblock \showarticletitle{Too human and not human enough: A grounded theory analysis of mental health harms from emotional dependence on the social chatbot Replika}.
\newblock \bibinfo{journal}{\emph{New Media \& Society}} \bibinfo{volume}{26}, \bibinfo{number}{10} (\bibinfo{year}{2024}), \bibinfo{pages}{5923--5941}.
\newblock
\showeprint{https://doi.org/10.1177/14614448221142007}
\href{https://doi.org/10.1177/14614448221142007}{doi:\nolinkurl{10.1177/14614448221142007}}


\bibitem[Li et~al\mbox{.}(2023)]%
        {chen2023nlpsuicide}
\bibfield{author}{\bibinfo{person}{Tim M~H Li}, \bibinfo{person}{Jie Chen}, \bibinfo{person}{Framenia O~C Law}, \bibinfo{person}{Chun-Tung Li}, \bibinfo{person}{Ngan~Yin Chan}, \bibinfo{person}{Joey W~Y Chan}, \bibinfo{person}{Steven W~H Chau}, \bibinfo{person}{Yaping Liu}, \bibinfo{person}{Shirley~Xin Li}, \bibinfo{person}{Jihui Zhang}, \bibinfo{person}{Kwong-Sak Leung}, {and} \bibinfo{person}{Yun-Kwok Wing}.} \bibinfo{year}{2023}\natexlab{}.
\newblock \showarticletitle{Detection of Suicidal Ideation in Clinical Interviews for Depression Using Natural Language Processing and Machine Learning: Cross-Sectional Study}.
\newblock \bibinfo{journal}{\emph{JMIR Med Inform}}  \bibinfo{volume}{11} (\bibinfo{date}{1 Dec} \bibinfo{year}{2023}), \bibinfo{pages}{e50221}.
\newblock
\showISSN{2291-9694}
\href{https://doi.org/10.2196/50221}{doi:\nolinkurl{10.2196/50221}}


\bibitem[Li et~al\mbox{.}(2025)]%
        {li2025counselbench}
\bibfield{author}{\bibinfo{person}{Yahan Li}, \bibinfo{person}{Jifan Yao}, \bibinfo{person}{John Bosco~S. Bunyi}, \bibinfo{person}{Adam~C. Frank}, \bibinfo{person}{Angel Hwang}, {and} \bibinfo{person}{Ruishan Liu}.} \bibinfo{year}{2025}\natexlab{}.
\newblock \bibinfo{title}{CounselBench: A Large-Scale Expert Evaluation and Adversarial Benchmarking of Large Language Models in Mental Health Question Answering}.
\newblock
\showeprint[arxiv]{2506.08584}~[cs.CL]
\urldef\tempurl%
\url{https://arxiv.org/abs/2506.08584}
\showURL{%
\tempurl}


\bibitem[Linden(2013)]%
        {linden2013iatrogenic}
\bibfield{author}{\bibinfo{person}{Michael Linden}.} \bibinfo{year}{2013}\natexlab{}.
\newblock \showarticletitle{How to define, find and classify side effects in psychotherapy: from unwanted events to adverse treatment reactions}.
\newblock \bibinfo{journal}{\emph{Clinical Psychology \& Psychotherapy}} \bibinfo{volume}{20}, \bibinfo{number}{4} (\bibinfo{year}{2013}), \bibinfo{pages}{286--296}.
\newblock
\href{https://doi.org/10.1002/cpp.1765}{doi:\nolinkurl{10.1002/cpp.1765}}


\bibitem[Liu et~al\mbox{.}(2023)]%
        {liu2023chatcounselor}
\bibfield{author}{\bibinfo{person}{June~M. Liu}, \bibinfo{person}{Donghao Li}, \bibinfo{person}{He Cao}, \bibinfo{person}{Tianhe Ren}, \bibinfo{person}{Zeyi Liao}, {and} \bibinfo{person}{Jiamin Wu}.} \bibinfo{year}{2023}\natexlab{}.
\newblock \bibinfo{title}{ChatCounselor: A Large Language Models for Mental Health Support}.
\newblock
\showeprint[arxiv]{2309.15461}~[cs.CL]
\urldef\tempurl%
\url{https://arxiv.org/abs/2309.15461}
\showURL{%
\tempurl}


\bibitem[Liu et~al\mbox{.}(2020)]%
        {liu2020passive}
\bibfield{author}{\bibinfo{person}{Richard~T. Liu}, \bibinfo{person}{Alexandra~H. Bettis}, {and} \bibinfo{person}{Taylor~A. Burke}.} \bibinfo{year}{2020}\natexlab{}.
\newblock \showarticletitle{Characterizing the phenomenology of passive suicidal ideation: a systematic review and meta-analysis of its prevalence, psychiatric comorbidity, correlates, and comparisons with active suicidal ideation}.
\newblock \bibinfo{journal}{\emph{Psychological Medicine}} \bibinfo{volume}{50}, \bibinfo{number}{3} (\bibinfo{year}{2020}), \bibinfo{pages}{367–383}.
\newblock
\href{https://doi.org/10.1017/S003329171900391X}{doi:\nolinkurl{10.1017/S003329171900391X}}


\bibitem[Manhas et~al\mbox{.}(2025)]%
        {care}
\bibfield{author}{\bibinfo{person}{Prabal Manhas}, \bibinfo{person}{Joydeep Chandra}, \bibinfo{person}{Ramanjot Kaur}, {and} \bibinfo{person}{Rashi Sahay}.} \bibinfo{year}{2025}\natexlab{}.
\newblock \showarticletitle{Care Compass: Monitoring made easy, Care made effective}. In \bibinfo{booktitle}{\emph{2025 International Conference on Intelligent and Innovative Technologies in Computing, Electrical and Electronics (IITCEE)}}. \bibinfo{pages}{1--7}.
\newblock
\href{https://doi.org/10.1109/IITCEE64140.2025.10915307}{doi:\nolinkurl{10.1109/IITCEE64140.2025.10915307}}


\bibitem[Menon(2024)]%
        {mental-health-mistral-7b}
\bibfield{author}{\bibinfo{person}{G.~R. Menon}.} \bibinfo{year}{2024}\natexlab{}.
\newblock \bibinfo{title}{mental-health-mistral-7b-instructv0.2-finetuned-V2}.
\newblock \bibinfo{howpublished}{\url{https://huggingface.co/GRMenon/mental-health-mistral-7b-instructv0.2-finetuned-V2}}.
\newblock


\bibitem[Miles et~al\mbox{.}(2018)]%
        {miles2014qualitative}
\bibfield{author}{\bibinfo{person}{Matthew~B. Miles}, \bibinfo{person}{A.~Michael Huberman}, {and} \bibinfo{person}{Johnny Salda{\~n}a}.} \bibinfo{year}{2018}\natexlab{}.
\newblock \bibinfo{booktitle}{\emph{Qualitative Data Analysis: A Methods Sourcebook} (\bibinfo{edition}{4th} ed.)}.
\newblock \bibinfo{publisher}{SAGE Publications, Inc}, \bibinfo{address}{Thousand Oaks, CA}.
\newblock
\showISBNx{978-1506353074}


\bibitem[Muran et~al\mbox{.}(2021)]%
        {muran2021rupture}
\bibfield{author}{\bibinfo{person}{J.~Christopher Muran}, \bibinfo{person}{Catherine~F. Eubanks}, {and} \bibinfo{person}{Lisa~Wallner Samstag}.} \bibinfo{year}{2021}\natexlab{}.
\newblock \showarticletitle{One more time with less jargon: An introduction to ``Rupture Repair in Practice''}.
\newblock \bibinfo{journal}{\emph{Journal of Clinical Psychology}} \bibinfo{volume}{77}, \bibinfo{number}{2} (\bibinfo{year}{2021}), \bibinfo{pages}{361--368}.
\newblock
\href{https://doi.org/10.1002/jclp.23105}{doi:\nolinkurl{10.1002/jclp.23105}}


\bibitem[Navneet et~al\mbox{.}(2026)]%
        {orches}
\bibfield{author}{\bibinfo{person}{Satyam~Kumar Navneet}, \bibinfo{person}{Joydeep Chandra}, {and} \bibinfo{person}{Yong Zhang}.} \bibinfo{year}{2026}\natexlab{}.
\newblock \bibinfo{title}{Orchestrating Attention: Bringing Harmony to the 'Chaos' of Neurodivergent Learning States}.
\newblock
\showeprint[arxiv]{2602.07865}~[cs.HC]
\urldef\tempurl%
\url{https://arxiv.org/abs/2602.07865}
\showURL{%
\tempurl}


\bibitem[Nolen-Hoeksema(2000)]%
        {nolen2000rumination}
\bibfield{author}{\bibinfo{person}{S Nolen-Hoeksema}.} \bibinfo{year}{2000}\natexlab{}.
\newblock \showarticletitle{The Role of Rumination in Depressive Disorders and Mixed Anxiety/Depressive Symptoms}.
\newblock \bibinfo{journal}{\emph{Journal of Abnormal Psychology}} \bibinfo{volume}{109}, \bibinfo{number}{3} (\bibinfo{year}{2000}), \bibinfo{pages}{504--511}.
\newblock
\href{https://doi.org/10.1037/0021-843X.109.3.504}{doi:\nolinkurl{10.1037/0021-843X.109.3.504}}


\bibitem[Ovsyannikova et~al\mbox{.}(2025)]%
        {ovsyannikova2025aicompassion}
\bibfield{author}{\bibinfo{person}{Daria Ovsyannikova}, \bibinfo{person}{Victor~O. de Mello}, {and} \bibinfo{person}{Michael Inzlicht}.} \bibinfo{year}{2025}\natexlab{}.
\newblock \showarticletitle{Third-party evaluators perceive AI as more compassionate than expert humans}.
\newblock \bibinfo{journal}{\emph{Communications Psychology}} \bibinfo{volume}{3}, \bibinfo{number}{1} (\bibinfo{year}{2025}), \bibinfo{pages}{4}.
\newblock
\href{https://doi.org/10.1038/s44271-024-00182-6}{doi:\nolinkurl{10.1038/s44271-024-00182-6}}


\bibitem[Park et~al\mbox{.}(2023)]%
        {park2023generative}
\bibfield{author}{\bibinfo{person}{Joon~Sung Park}, \bibinfo{person}{Joseph O'Brien}, \bibinfo{person}{Carrie~Jun Cai}, \bibinfo{person}{Meredith~Ringel Morris}, \bibinfo{person}{Percy Liang}, {and} \bibinfo{person}{Michael~S. Bernstein}.} \bibinfo{year}{2023}\natexlab{}.
\newblock \showarticletitle{Generative Agents: Interactive Simulacra of Human Behavior}. In \bibinfo{booktitle}{\emph{Proceedings of the 36th Annual ACM Symposium on User Interface Software and Technology}} (San Francisco, CA, USA) \emph{(\bibinfo{series}{UIST '23})}. \bibinfo{publisher}{Association for Computing Machinery}, \bibinfo{address}{New York, NY, USA}, Article \bibinfo{articleno}{2}, \bibinfo{numpages}{22}~pages.
\newblock
\showISBNx{9798400701320}
\href{https://doi.org/10.1145/3586183.3606763}{doi:\nolinkurl{10.1145/3586183.3606763}}


\bibitem[Parry et~al\mbox{.}(2016)]%
        {parry2016iatrogenic}
\bibfield{author}{\bibinfo{person}{Glenys~D. Parry}, \bibinfo{person}{Mike~J. Crawford}, {and} \bibinfo{person}{Conor Duggan}.} \bibinfo{year}{2016}\natexlab{}.
\newblock \showarticletitle{Iatrogenic harm from psychological therapies – time to moveon}.
\newblock \bibinfo{journal}{\emph{British Journal of Psychiatry}} \bibinfo{volume}{208}, \bibinfo{number}{3} (\bibinfo{year}{2016}), \bibinfo{pages}{210–212}.
\newblock
\href{https://doi.org/10.1192/bjp.bp.115.163618}{doi:\nolinkurl{10.1192/bjp.bp.115.163618}}


\bibitem[Pentina et~al\mbox{.}(2023)]%
        {pentina2023replika}
\bibfield{author}{\bibinfo{person}{Iryna Pentina}, \bibinfo{person}{Tyler Hancock}, {and} \bibinfo{person}{Tianling Xie}.} \bibinfo{year}{2023}\natexlab{}.
\newblock \showarticletitle{Exploring relationship development with social chatbots: A mixed-method study of replika}.
\newblock \bibinfo{journal}{\emph{Comput. Hum. Behav.}} \bibinfo{volume}{140}, \bibinfo{number}{C} (\bibinfo{date}{March} \bibinfo{year}{2023}), \bibinfo{numpages}{15}~pages.
\newblock
\showISSN{0747-5632}
\href{https://doi.org/10.1016/j.chb.2022.107600}{doi:\nolinkurl{10.1016/j.chb.2022.107600}}


\bibitem[Qiu et~al\mbox{.}(2025)]%
        {qiu2025emoagent}
\bibfield{author}{\bibinfo{person}{Jiahao Qiu}, \bibinfo{person}{Yinghui He}, \bibinfo{person}{Xinzhe Juan}, \bibinfo{person}{Yimin Wang}, \bibinfo{person}{Yuhan Liu}, \bibinfo{person}{Zixin Yao}, \bibinfo{person}{Yue Wu}, \bibinfo{person}{Xun Jiang}, \bibinfo{person}{Ling Yang}, {and} \bibinfo{person}{Mengdi Wang}.} \bibinfo{year}{2025}\natexlab{}.
\newblock \bibinfo{title}{EmoAgent: Assessing and Safeguarding Human-AI Interaction for Mental Health Safety}.
\newblock
\showeprint[arxiv]{2504.09689}~[cs.AI]
\urldef\tempurl%
\url{https://arxiv.org/abs/2504.09689}
\showURL{%
\tempurl}


\bibitem[Safran et~al\mbox{.}(2011)]%
        {safran2011therapeutic}
\bibfield{author}{\bibinfo{person}{Jeremy~D. Safran}, \bibinfo{person}{J.~Christopher Muran}, {and} \bibinfo{person}{Catherine Eubanks-Carter}.} \bibinfo{year}{2011}\natexlab{}.
\newblock \showarticletitle{Repairing alliance ruptures}.
\newblock \bibinfo{journal}{\emph{Psychotherapy}} \bibinfo{volume}{48}, \bibinfo{number}{1} (\bibinfo{year}{2011}), \bibinfo{pages}{80--87}.
\newblock
\href{https://doi.org/10.1037/a0022140}{doi:\nolinkurl{10.1037/a0022140}}


\bibitem[Sedlakova and Trachsel(2023)]%
        {rubeis2022cai}
\bibfield{author}{\bibinfo{person}{Jana Sedlakova} {and} \bibinfo{person}{Manuel Trachsel}.} \bibinfo{year}{2023}\natexlab{}.
\newblock \showarticletitle{Conversational Artificial Intelligence in Psychotherapy: A New Therapeutic Tool or Agent?}
\newblock \bibinfo{journal}{\emph{The American Journal of Bioethics}} \bibinfo{volume}{23}, \bibinfo{number}{5} (\bibinfo{year}{2023}), \bibinfo{pages}{4--13}.
\newblock
\href{https://doi.org/10.1080/15265161.2022.2048739}{doi:\nolinkurl{10.1080/15265161.2022.2048739}}


\bibitem[Team(2021)]%
        {minilm}
\bibfield{author}{\bibinfo{person}{MiniLM Team}.} \bibinfo{year}{2021}\natexlab{}.
\newblock \bibinfo{title}{all-MiniLM-L6-v2: Sentence-Transformers Model}.
\newblock \bibinfo{howpublished}{\url{https://huggingface.co/sentence-transformers/all-MiniLM-L6-v2}}.
\newblock


\bibitem[Touvron et~al\mbox{.}(2023)]%
        {llama2}
\bibfield{author}{\bibinfo{person}{Hugo Touvron}, \bibinfo{person}{Louis Martin}, \bibinfo{person}{Kevin Stone}, \bibinfo{person}{Peter Albert}, \bibinfo{person}{Amjad Almahairi}, \bibinfo{person}{Yasmine Babaei}, \bibinfo{person}{Nikolay Bashlykov}, \bibinfo{person}{Soumya Batra}, \bibinfo{person}{Prajjwal Bhargava}, \bibinfo{person}{Shruti Bhosale}, \bibinfo{person}{Dan Bikel}, \bibinfo{person}{Lukas Blecher}, \bibinfo{person}{Cristian~Canton Ferrer}, \bibinfo{person}{Moya Chen}, \bibinfo{person}{Guillem Cucurull}, \bibinfo{person}{David Esiobu}, \bibinfo{person}{Jude Fernandes}, \bibinfo{person}{Jeremy Fu}, \bibinfo{person}{Wenyin Fu}, \bibinfo{person}{Brian Fuller}, \bibinfo{person}{Cynthia Gao}, \bibinfo{person}{Vedanuj Goswami}, \bibinfo{person}{Naman Goyal}, \bibinfo{person}{Anthony Hartshorn}, \bibinfo{person}{Saghar Hosseini}, \bibinfo{person}{Rui Hou}, \bibinfo{person}{Hakan Inan}, \bibinfo{person}{Marcin Kardas}, \bibinfo{person}{Viktor Kerkez}, \bibinfo{person}{Madian Khabsa},
  \bibinfo{person}{Isabel Kloumann}, \bibinfo{person}{Artem Korenev}, \bibinfo{person}{Punit~Singh Koura}, \bibinfo{person}{Marie-Anne Lachaux}, \bibinfo{person}{Thibaut Lavril}, \bibinfo{person}{Jenya Lee}, \bibinfo{person}{Diana Liskovich}, \bibinfo{person}{Yinghai Lu}, \bibinfo{person}{Yuning Mao}, \bibinfo{person}{Xavier Martinet}, \bibinfo{person}{Todor Mihaylov}, \bibinfo{person}{Pushkar Mishra}, \bibinfo{person}{Igor Molybog}, \bibinfo{person}{Yixin Nie}, \bibinfo{person}{Andrew Poulton}, \bibinfo{person}{Jeremy Reizenstein}, \bibinfo{person}{Rashi Rungta}, \bibinfo{person}{Kalyan Saladi}, \bibinfo{person}{Alan Schelten}, \bibinfo{person}{Ruan Silva}, \bibinfo{person}{Eric~Michael Smith}, \bibinfo{person}{Ranjan Subramanian}, \bibinfo{person}{Xiaoqing~Ellen Tan}, \bibinfo{person}{Binh Tang}, \bibinfo{person}{Ross Taylor}, \bibinfo{person}{Adina Williams}, \bibinfo{person}{Jian~Xiang Kuan}, \bibinfo{person}{Puxin Xu}, \bibinfo{person}{Zheng Yan}, \bibinfo{person}{Iliyan Zarov}, \bibinfo{person}{Yuchen
  Zhang}, \bibinfo{person}{Angela Fan}, \bibinfo{person}{Melanie Kambadur}, \bibinfo{person}{Sharan Narang}, \bibinfo{person}{Aurelien Rodriguez}, \bibinfo{person}{Robert Stojnic}, \bibinfo{person}{Sergey Edunov}, {and} \bibinfo{person}{Thomas Scialom}.} \bibinfo{year}{2023}\natexlab{}.
\newblock \bibinfo{title}{Llama 2: Open Foundation and Fine-Tuned Chat Models}.
\newblock
\showeprint[arxiv]{2307.09288}~[cs.CL]
\urldef\tempurl%
\url{https://arxiv.org/abs/2307.09288}
\showURL{%
\tempurl}


\bibitem[{UNESCO}(2025)]%
        {unesco2025chatbot}
\bibfield{author}{\bibinfo{person}{{UNESCO}}.} \bibinfo{year}{2025}\natexlab{}.
\newblock \showarticletitle{Ghost in the Chatbot: The Perils of Parasocial Attachment}.
\newblock \bibinfo{journal}{\emph{UNESCO Articles}} (\bibinfo{date}{October} \bibinfo{year}{2025}).
\newblock
\urldef\tempurl%
\url{https://www.unesco.org/en/articles/ghost-chatbot-perils-parasocial-attachment}
\showURL{%
\tempurl}


\bibitem[Wang et~al\mbox{.}(2025)]%
        {sharma2025crbot}
\bibfield{author}{\bibinfo{person}{Yinzhou Wang}, \bibinfo{person}{Yimeng Wang}, \bibinfo{person}{Ye Xiao}, \bibinfo{person}{Liabette Escamilla}, \bibinfo{person}{Bianca Augustine}, \bibinfo{person}{Kelly Crace}, \bibinfo{person}{Gang Zhou}, {and} \bibinfo{person}{Yixuan Zhang}.} \bibinfo{year}{2025}\natexlab{}.
\newblock \bibinfo{title}{Evaluating an LLM-Powered Chatbot for Cognitive Restructuring: Insights from Mental Health Professionals}.
\newblock
\showeprint[arxiv]{2501.15599}~[cs.HC]
\urldef\tempurl%
\url{https://arxiv.org/abs/2501.15599}
\showURL{%
\tempurl}


\bibitem[Xu et~al\mbox{.}(2025)]%
        {xu2025dta}
\bibfield{author}{\bibinfo{person}{Z. Xu}, \bibinfo{person}{Y. Lee}, \bibinfo{person}{K. Stasiak}, \bibinfo{person}{J. Warren}, {and} \bibinfo{person}{D. Lottridge}.} \bibinfo{year}{2025}\natexlab{}.
\newblock \showarticletitle{The Digital Therapeutic Alliance With Mental Health Chatbots: Diary Study and Thematic Analysis}.
\newblock \bibinfo{journal}{\emph{JMIR Mental Health}}  \bibinfo{volume}{12} (\bibinfo{year}{2025}), \bibinfo{pages}{e76642}.
\newblock
\href{https://doi.org/10.2196/76642}{doi:\nolinkurl{10.2196/76642}}


\bibitem[Yang(2024)]%
        {nbcnews2024characterai}
\bibfield{author}{\bibinfo{person}{Angela Yang}.} \bibinfo{year}{2024}\natexlab{}.
\newblock \bibinfo{title}{Lawsuit claims Character.AI is responsible for teen's suicide}.
\newblock \bibinfo{howpublished}{NBC News}.
\newblock
\urldef\tempurl%
\url{https://www.nbcnews.com/tech/characterai-lawsuit-florida-teen-death-rcna176791}
\showURL{%
\tempurl}


\bibitem[Yang et~al\mbox{.}(2024)]%
        {yang2023mentalllama}
\bibfield{author}{\bibinfo{person}{Kailai Yang}, \bibinfo{person}{Tianlin Zhang}, \bibinfo{person}{Ziyan Kuang}, \bibinfo{person}{Qianqian Xie}, \bibinfo{person}{Jimin Huang}, {and} \bibinfo{person}{Sophia Ananiadou}.} \bibinfo{year}{2024}\natexlab{}.
\newblock \showarticletitle{MentaLLaMA: Interpretable Mental Health Analysis on Social Media with Large Language Models}. In \bibinfo{booktitle}{\emph{Proceedings of the ACM Web Conference 2024}} \emph{(\bibinfo{series}{WWW ’24})}. \bibinfo{publisher}{ACM}, \bibinfo{pages}{4489–4500}.
\newblock
\href{https://doi.org/10.1145/3589334.3648137}{doi:\nolinkurl{10.1145/3589334.3648137}}


\bibitem[Zhai et~al\mbox{.}(2025)]%
        {peng2025dependency}
\bibfield{author}{\bibinfo{person}{Na Zhai}, \bibinfo{person}{Xiaomei Ma}, {and} \bibinfo{person}{Xiaojun Ding}.} \bibinfo{year}{2025}\natexlab{}.
\newblock \showarticletitle{Unpacking AI Chatbot Dependency: A Dual-Path Model of Cognitive and Affective Mechanisms}.
\newblock \bibinfo{journal}{\emph{Information}} \bibinfo{volume}{16}, \bibinfo{number}{12} (\bibinfo{year}{2025}), \bibinfo{pages}{1025}.
\newblock
\href{https://doi.org/10.3390/info16121025}{doi:\nolinkurl{10.3390/info16121025}}


\bibitem[Zhang et~al\mbox{.}(2025)]%
        {zhang2025chatbotloneliness}
\bibfield{author}{\bibinfo{person}{Yutong Zhang}, \bibinfo{person}{Dora Zhao}, \bibinfo{person}{Jeffrey~T. Hancock}, \bibinfo{person}{Robert Kraut}, {and} \bibinfo{person}{Diyi Yang}.} \bibinfo{year}{2025}\natexlab{}.
\newblock \bibinfo{title}{The Rise of AI Companions: How Human-Chatbot Relationships Influence Well-Being}.
\newblock
\showeprint[arxiv]{2506.12605}~[cs.HC]
\urldef\tempurl%
\url{https://arxiv.org/abs/2506.12605}
\showURL{%
\tempurl}


\end{thebibliography}

\end{document}